\documentclass[pdflatex,sn-mathphys-num]{sn-jnl}

\usepackage{graphicx}%
\usepackage{multirow}%
\usepackage{amsmath,amssymb,amsfonts}%
\usepackage{amsthm}%
\usepackage{mathrsfs}%
\usepackage[title]{appendix}%
\usepackage{xcolor}%
\usepackage{textcomp}%
\usepackage{manyfoot}%
\usepackage{booktabs}%
\usepackage{algorithm}%
\usepackage{algorithmicx}%
\usepackage{algpseudocode}%
\usepackage{listings}%

\theoremstyle{thmstyleone}%
\theoremstyle{thmstyletwo}%

\theoremstyle{thmstylethree}%

\begin{document}

\title[Learning Time-Varying Connectomes for Stable Behavior Across Development]{Learning Time-Varying Connectomes for Stable Behavior Across Development}


\author*[1]{\fnm{Christof} \sur{Fehrman}}\email{christof.fehrman@duke.edu}

\author[1]{\fnm{Jacob J.} \sur{Morra}}\email{jacob.morra@duke.edu}

\author[1]{\fnm{Eva A.} \sur{Naumann}}\email{eva.naumann@duke.edu}

\affil*[1]{\orgdiv{Neurobiology}, \orgname{Duke University}, \orgaddress{ \city{Durham}, \postcode{27710}, \state{NC}, \country{USA}}}


\abstract{Neural circuits must maintain functionality to drive effective behavior while their size, topology, geometry, and intrinsic dynamics change across development. To model developing visuomotor circuits and behavior, we investigated whether a single compact generative encoding could produce functional recurrent controllers across such changes. Building on the HyperNEAT algorithm, we represented this program as a compositional pattern-producing network that received developmental time as an input to generate age-specific connectivity. The developmental substrate prescribed neuronal birth, position, sensory organization, and intrinsic time constants. We tested this approach using two geometrically distinct developmental substrates and two visuomotor tasks: optomotor responses and phototaxis. Controllers optimized at only the initial or mature substrate generally failed when evaluated at other ages. In contrast, controllers optimized across multiple developmental snapshots maintained substantially higher and more consistent performance across trained and unseen ages. Optimization at only three developmental snapshots was often sufficient to produce effective behavior across the tested developmental interval. The resulting connectomes changed substantially with age and did not converge on a single trajectory of sparsity or connection-weight statistics. Controllers also differed in whether they relied on neurons born later in development. These results demonstrate that a shared spatiotemporal time-varying connectivity program can produce recurrent neural networks that preserve behavior across substantial developmental changes. Together, these results provide a framework for studying how functional stability can emerge without requiring structural or dynamical stability.}

\keywords{Developmental neuroevolution, Embodied agents, Evolutionary algorithms, Recurrent neural networks}



\maketitle

\section{Introduction}\label{sec1}

Neural development involves the birth and maturation of neurons, the formation and elimination of synapses, changes in neuronal position and morphology, and continuing refinement of intrinsic and synaptic properties \cite{goodhill2018theoretical, richter2017understanding}. Since an animal cannot wait until it reaches maturity before interacting successfully with its environment, the nervous system must produce behavior that supports survival throughout these changes. Neural development should therefore be understood not merely as the construction of a mature nervous system, but as a process that sustains useful computation along the path to maturity.

Preserving behavioral function across development does not require the underlying circuit to remain unchanged. For example, phase relationships in the lobster pyloric motor rhythm remain similar between juvenile and adult animals despite pronounced changes in neural and body size \cite{bucher2005animal}, and neurons established during embryogenesis can retain important locomotor functions across life stages despite extensive structural remodeling \cite{padmanabhan2026embryonic}. Similar network outputs can arise from substantially different combinations of intrinsic and synaptic parameters \cite{prinz2004similar}, and developing circuits can recruit different combinations of homeostatic mechanisms across cell types, perturbations, and developmental stages \cite{tien2018homeostatic}. The detailed expression of a behavior may also change as the nervous system and body develop, even when the behavior continues to serve the same function. How developing nervous systems preserve behavioral function despite substantial changes in connectivity and neural dynamics remains unclear.

Neural development has received substantial computational attention, but existing approaches typically capture only part of this problem. Many models describe developmental changes in synaptic weights, intrinsic excitability, or neural activity while retaining a fixed set of neurons and allowable connections \cite{zavitz2025modeling}. Such models can reveal how basic circuit computation changes during development, but cannot directly represent neurogenesis, changing neural geometry, or changes in the fundamental network topology. Other methods explicitly enlarge artificial neural networks, often by adding neurons or layers through a small number of discrete transformations. These transformations may preserve a feedforward network’s function \cite{chen2015net2net}, increase its capacity during training \cite{fehring2025growing}, or allow it to acquire new information without erasing existing representations \cite{draelos2017neurogenesis}. Although these methods permit structural change, they generally represent development as a small number of engineered transitions within layered architectures rather than as progressive change in the substrate of a neural circuit.

Developmental encodings provide a more flexible alternative by specifying local or generative rules through which network structure emerges. Recent neural developmental programs can add neurons, differentiate cell states, and construct networks with changing topologies \cite{najarro2023towards, nisioti2024growing}. These methods produce substantial structural change, but performance is evaluated only after circuit growth with the intermediate developmental controllers not required to produce effective behavior. Related work extends structural plasticity into the agent’s lifetime but focuses on experience- and reward-dependent adaptation rather than preservation of a selected function across a prescribed developmental trajectory \cite{plantec2024evolving}. Together, these approaches leave open whether a controller can sustain high behavioral performance as its neural substrate changes over developmental time. Maintaining function throughout structural development is especially challenging for recurrent controllers. In a layered feedforward network, additional units can sometimes be introduced through algebraic transformations that preserve a static input–output mapping \cite{chen2015net2net}. In recurrent neural networks, changes in recurrent connectivity and neuronal time constants can alter transient response times, equilibrium stability, oscillatory regimes, and attractor dynamics \cite{beer1995dynamics,krauss2019weight}. Because recurrent controllers operate within a closed sensorimotor loop, such internal changes may also reshape the coupled dynamics of the controller, body, and environment \cite{aguilera2013situated}. Adding neurons that are synaptically integrated into the recurrent network necessarily changes its connectivity and can therefore alter the dynamics of the coupled system. Functional continuity therefore cannot be inferred from architectural similarity or parameter preservation alone and must instead be evaluated at the level of behavior.

Representing an entire developmental trajectory of connectomes creates an additional challenge. One approach would be to specify every synapse independently at every developmental age, but such a representation would be inefficient and a poor abstraction of biological development. The information capacity of the genome is insufficient to enumerate the full wiring diagram of a complex nervous system; instead, genomes are thought to specify compact rules that guide circuit formation \cite{zador2019critique}. This creates a ``genomic bottleneck'' through which the information required to construct much larger neural systems must pass \cite{shuvaev2024encoding}. Hypernetworks provide an analogous computational strategy in which one network generates parameters for another, typically larger network \cite{ha2016hypernetworks, chauhan2024brief}. Such a generator can reuse shared structure rather than independently encoding every connection in every network it produces. This perspective motivates the more specific hypothesis that developing connectomes possess compressible spatiotemporal structure that can be represented directly as a function of neural geometry and developmental time. Because each developmental stage modifies the preceding network, both connectome structure and the neural activity it supports are expected to exhibit temporal continuity rather than vary independently across ages \cite{witvliet2021connectomes,majnik2025longitudinal}. This continuity does not require connections, intrinsic dynamics, or circuit roles to remain invariant. Rather, it implies that age-specific connectomes share sufficient structure to be parameterized by a common conditional generator. The persistence of neuronal function across extensive developmental remodeling provides biological motivation for this possibility \cite{padmanabhan2026embryonic}. 

Hypercube-based NeuroEvolution of Augmenting Topologies (HyperNEAT) provides a natural framework for testing this hypothesis \cite{stanley2009hypercube}. HyperNEAT evolves compositional pattern-producing networks (CPPNs) that generate neural connectivity as a function of substrate geometry, allowing a compact encoding to parameterize networks containing many connections. CPPNs were originally proposed as an abstraction of biological development, in which a compact genotype generates spatial structure without explicitly simulating cell-by-cell growth \citep{stanley2007compositional}. In previous HyperNEAT work, development was represented by applying the same CPPN to substrates of increasing neuronal density \citep{stanley2009hypercube}. Density scaling demonstrates how a compact encoding can elaborate a larger network, but it is not by itself a realistic model of neural development. It omits the coordinated changes described above, including changes in neuronal position, sensory organization, intrinsic cellular dynamics, and connectivity. We therefore ask whether HyperNEAT's compact generative encoding can support developmental trajectories that incorporate these broader changes. CPPNs can also receive contextual variables as inputs to generate controllers appropriate to different conditions. Morphology-conditioned HyperNEAT provides a related precedent. \citet{risi2013confronting} conditioned a CPPN on quadruped leg length and evolved controllers that generalized to intermediate leg lengths not encountered during optimization. Their CPPN generated morphology-dependent connection weights, neuronal biases, and time constants. However, the controller substrate itself remained fixed across robot morphologies: the same neurons, neural coordinates, sensory and motor organization, and set of possible connections were retained. Neural development presents a different problem because the neural substrate itself changes over developmental time. Whether a shared generator can maintain function while the composition, geometry, and intrinsic dynamics of that substrate change therefore remains unknown.

Here, we formulate developmental functional stability as a trajectory-level optimization problem. We evolve a compact connectivity generator that receives neural geometry and developmental time as inputs and generates the connectivity of a recurrent controller for the substrate at any point of development (Fig. 1). The developmental environment prescribes which neurons are present, where they are located, and how their intrinsic dynamics change, while the same evolved generator parameterizes connectivity throughout the developmental trajectory. Rather than evaluating only the controller produced at maturity, we optimize the generator at multiple developmental snapshots and evaluate its performance across the full developmental interval. We define developmental functional stability as sustained high task performance across age, without requiring invariant network parameters, neural dynamics, or behavioral trajectories.

Because the regularities available to HyperNEAT depend on how the substrate is geometrically represented \cite{clune2009sensitivity}, evidence from a single substrate would not establish that the approach extends beyond one favorable geometry. We therefore evaluate the method using two geometrically distinct models of neural development and two visuomotor behaviors. One substrate (Fish) is inspired by the organization and growth of a developing fish visuomotor system \cite{naumann2016whole}, whereas the second (Ring) uses an abstract ring geometry with a substantially different pattern of neural growth (see Methods). Each substrate is evaluated on optomotor response (OMR), a visually guided response to global motion \cite{naumann2016whole}, and phototaxis, navigation toward a light source \cite{wolf2017sensorimotor}. Aside from the definitions and constraints required by each substrate and task, the same evolutionary procedure and hyperparameters are used across the four substrate–task combinations. Across these conditions, we find that optimizing a shared connectivity generator at only a few developmental snapshots is typically sufficient to maintain high task performance throughout development, including at ages not encountered during optimization.
\begin{figure}[h]
\centering
\includegraphics[]{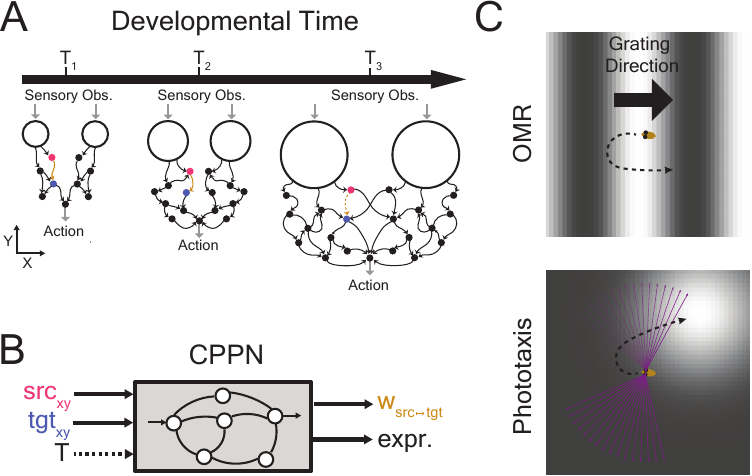}
\caption{Parameterizing the connectome of a developing visuomotor system. (A) Three developmental snapshots illustrate the challenge of maintaining function as the neural substrate changes. The system must use sensory observations to produce survival-relevant actions while accommodating changes in network topology (e.g., neurogenesis) through local synaptic modifications without disrupting behavior. (B) A compositional pattern-producing network (CPPN) parameterizes the connectome. In HyperNEAT, source (src) and target (tgt) neuron coordinates determine directed synaptic weights, while an optional expression gate (expr.) promotes modularity \citep{verbancsics2011constraining}. Our developmentally aware modification adds developmental time ($T$) as a CPPN input, enabling time-varying generative rules. (C) CPPNs are optimized for optomotor response (OMR; top), in which the agent follows moving gratings, or phototaxis (bottom), in which it moves toward a stationary light source (purple rays indicate the retinal detection radius; see Methods). Although training uses only the three illustrated snapshots, performance is evaluated across the developmental trajectory. The example synaptic weight (yellow) between a source (red) and target (blue) in A may vary with substrate changes across the first two snapshots while in the third, the expression gate falls below threshold, clamping the weight to zero (dotted line). The objective is to maintain functionally stable behavior throughout development despite changes in the underlying connectome.}\label{fig1}
\end{figure}

\section{Results}\label{sec2}
\subsection{Five optimization conditions across neural development}
Because HyperNEAT has been proposed as a model of neural development, we first asked whether the standard approach could accommodate the more extensive substrate growth considered here. In developmental applications of HyperNEAT, a CPPN is commonly optimized on a smaller neural substrate and then queried on a larger substrate. Our HyperNEAT condition was optimized only on the initial substrate at $T = 0.0$ and subsequently evaluated throughout development. Any loss of performance could therefore arise either because the generated connectivity failed to generalize across development or because transferring from the smallest substrate to larger substrates was particularly difficult. To distinguish these possibilities, we included a ``reverse'' HyperNEAT condition (Rev. HyperNEAT) which was optimized only on the largest, fully developed substrate at $T = 1.0$ and then evaluated at earlier developmental times. Both endpoint-trained conditions therefore had to generalize away from a single developmental snapshot, but in opposite directions.

We compared these endpoint-trained controllers with three developmentally aware conditions. Dev. Aware–3pts, Dev. Aware–5pts, and Dev. Aware–15pts received developmental time as an additional CPPN input and were optimized for mean fitness across three, five, or fifteen evenly spaced developmental snapshots, respectively. Including multiple snapshots transformed evaluation at ages between the sampled points from endpoint extrapolation into interpolation within the training interval. We varied the number of snapshots for two reasons. First, it was unclear how few developmental observations would be sufficient to produce stable controllers or whether progressively denser sampling would yield diminishing improvements. Second, although additional snapshots provide denser coverage of development, they also impose a more demanding joint optimization problem because a single CPPN must perform well across an increasing number of substrates. The three conditions allowed us to examine this tradeoff between developmental sampling and optimization difficulty.

Five independent evolutionary runs were performed for each of the five conditions in every substrate–task combination, and all runs were optimized for 200 generations. Rev. HyperNEAT attained the highest training fitness in every substrate–task combination (Fig. 2A). This advantage was consistent with its optimization on the fully developed substrate, which contained the greatest number of neurons and therefore had the highest capacity. In contrast, HyperNEAT was optimized on the smaller initial substrate, and the developmentally aware conditions had to maximize mean performance across multiple substrates rather than at a single developmental endpoint. Despite this more demanding objective, the final training fitnesses of the developmentally aware conditions were generally comparable to those of HyperNEAT, with their relative ordering varying across substrate–task combinations (Fig. 2A,B). Training outcomes also depended on substrate geometry: evolutionary runs on the Ring substrate showed substantially less between-run variation than those on the Fish substrate for both tasks (Fig. 2A).

For subsequent analyses of individual controllers, we selected the evolutionary run from each condition with the highest mean training fitness over the final 10 generations (Fig. 2B). Selection was based only on the training objective and was therefore independent of performance at developmental times not encountered during optimization. We refer to these as the “best seeds” in subsequent analyses.

\subsection{Developmentally aware optimization maintained high fitness across trained and unseen ages}
We next evaluated every evolved controller at 11 evenly spaced developmental times from $T = 0.0$  to $T = 1.0$. At each developmental time, fitness was measured across 20 testing trials with random initializations and stimulus sequences not encountered during optimization. Fitness across this interval provides a trajectory-level measure of functional stability: a controller is stable when it continues to produce high task performance as the neural substrate develops.

Controllers optimized at a single developmental endpoint generally failed to transfer away from that endpoint. In the OMR task, HyperNEAT controllers performed well near $T = 0.0$ but typically declined sharply as development progressed, whereas Rev. HyperNEAT controllers performed well near $T = 1.0$ and declined toward earlier ages (Fig. 2C). One HyperNEAT evolutionary run showed partial stability across OMR development, but this behavior was not representative of the condition and remained less consistent than that of the developmentally aware conditions (Fig. 2C). The same endpoint dependence was apparent in phototaxis, where HyperNEAT and Rev. HyperNEAT controllers generally lost performance outside the age on which they had been optimized (Fig. 2D).

In contrast, controllers optimized across development maintained substantially higher and more consistent fitness over the developmental interval (Fig. 2C,D). This pattern was observed across both substrate geometries and both visuomotor tasks. Increasing the number of training snapshots generally produced more consistent developmental trajectories, but notably, even Dev. Aware–3pts controllers could interpolate across the intervals separating their training snapshots. This effect was especially pronounced for OMR on the Ring substrate, where all five Dev. Aware–3pts runs produced stable fitness trajectories despite having been optimized at only three developmental times (Fig. 2C). 

Developmental fitness was qualitatively more stable for OMR than for phototaxis (Fig. 2C,D). Nevertheless, optimizing a shared generator at a limited set of developmental snapshots was generally sufficient to preserve high performance throughout the developmental trajectory.

\begin{figure}[h]
\centering
\includegraphics[width=0.8\textwidth]{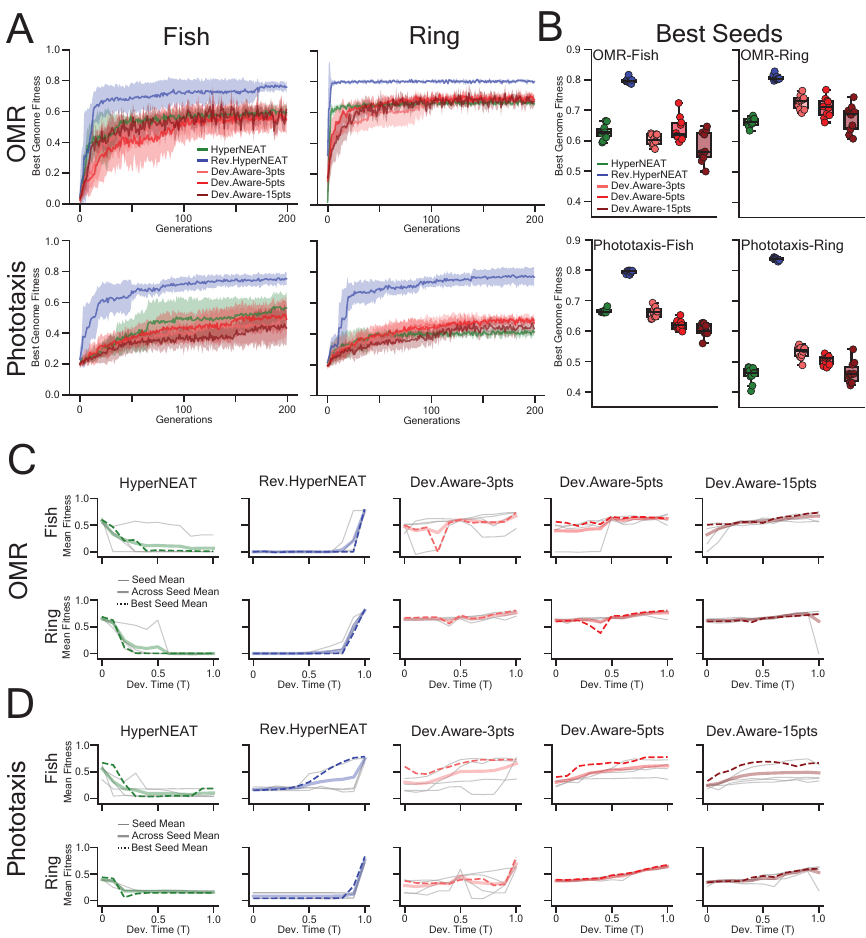}
\caption{Results of Task Optimization and Fitness Stability. (A) Fitness trajectories across training conditions and substrate–task pairs. Five evolutionary runs with different random seeds were performed per condition; bold lines show mean fitness across seeds, with standard deviations overlaid. Rev. HyperNEAT achieved the highest fitness on both substrates, while the Ring substrate showed lower variability across seeds. (B) Fitness over the final 10 generations for the best seed in each condition, selected by mean fitness over those generations. (C) Developmental robustness of OMR-trained networks, evaluated over 11 evenly spaced developmental times, most unseen during training. Gray lines show mean fitness across 20 test trials per seed and time point; thick transparent lines show condition means, and dotted lines indicate the best seed's fitness. HyperNEAT and Rev. HyperNEAT generally failed outside their optimized developmental times, whereas developmentally aware networks maintained more stable fitness. Notably, the highest training fitness rarely corresponded to the most fragile network. All five Ring seeds trained at three developmental time points (3pts) remained stable across development. (D) The same analysis for phototaxis reveals similar trends despite the task's greater difficulty, attributable to substantial changes in sensory structure (see Methods).}\label{fig2}
\end{figure}

\subsection{Developmentally optimized controllers produced consistent sensorimotor behavior at unseen ages}
High fitness does not require the neural activity or precise movement trajectory of an agent to remain identical throughout development. We therefore examined whether the training-selected controllers nevertheless produced qualitatively consistent behavior at different developmental ages. Each controller was evaluated in a fixed trial for values of $T$ not used in any condition’s optimization objective with the same stimulus statistics used in training (see Methods).

For OMR, behavior was visualized as the angular error between the agent’s orientation and the direction of grating motion. Developmentally aware controllers generally reduced this error following each change in grating direction and produced similar response patterns at all three developmental times (Fig. 3A). By comparison, the endpoint-trained controllers did not transfer their learned behaviors to the unseen ages. In the Fish substrate, the HyperNEAT controller was largely unresponsive to changes in grating direction, whereas the Rev. HyperNEAT controller frequently rotated continuously (Fig. 3A). The corresponding Ring controllers showed the opposite characteristic failures where HyperNEAT produced persistent rotation and Rev. HyperNEAT showed little or no orientation response (Fig. 3A). Although orientation error alone would not distinguish directed swimming from turning without forward movement, inspection of the underlying trajectories did not reveal controllers that achieved low angular error with stationary turning.

For phototaxis, behavior was visualized using the distance between the agent and the center of the light emitter. The developmentally aware controllers generally approached the emitter after its position changed and retained similar response patterns across the three unseen developmental times (Fig. 3B). The endpoint-trained controllers instead displayed age-dependent failures. On the Fish substrate, HyperNEAT showed some appropriate behavior at $T = 0.2$ but moved away from the emitter at later ages, while Rev. HyperNEAT responded at $T = 0.6$ and $T = 0.8$ but was largely unresponsive at $T = 0.2$ (Fig. 3B). On the Ring substrate, HyperNEAT moved away from the emitter at $T = 0.2$ and was largely unresponsive at the later ages, whereas Rev. HyperNEAT moved away from the emitter at all three ages (Fig. 3B). Several developmentally aware phototaxis controllers, particularly on the Ring substrate, produced oscillatory distance trajectories (Fig. 3B). Examination of the agents’ movements showed that these oscillations arose from repeated circular trajectories of different radii. We hypothesize these movements constituted an active scanning strategy since the location of the emitter would change at random intervals (see Methods). 

These example trajectories show that the stable fitness of developmentally aware controllers produced qualitatively coherent visuomotor behavior rather than unrelated high-performing trajectories at different ages.

\begin{figure}[h]
\centering
\includegraphics[width=0.8\textwidth]{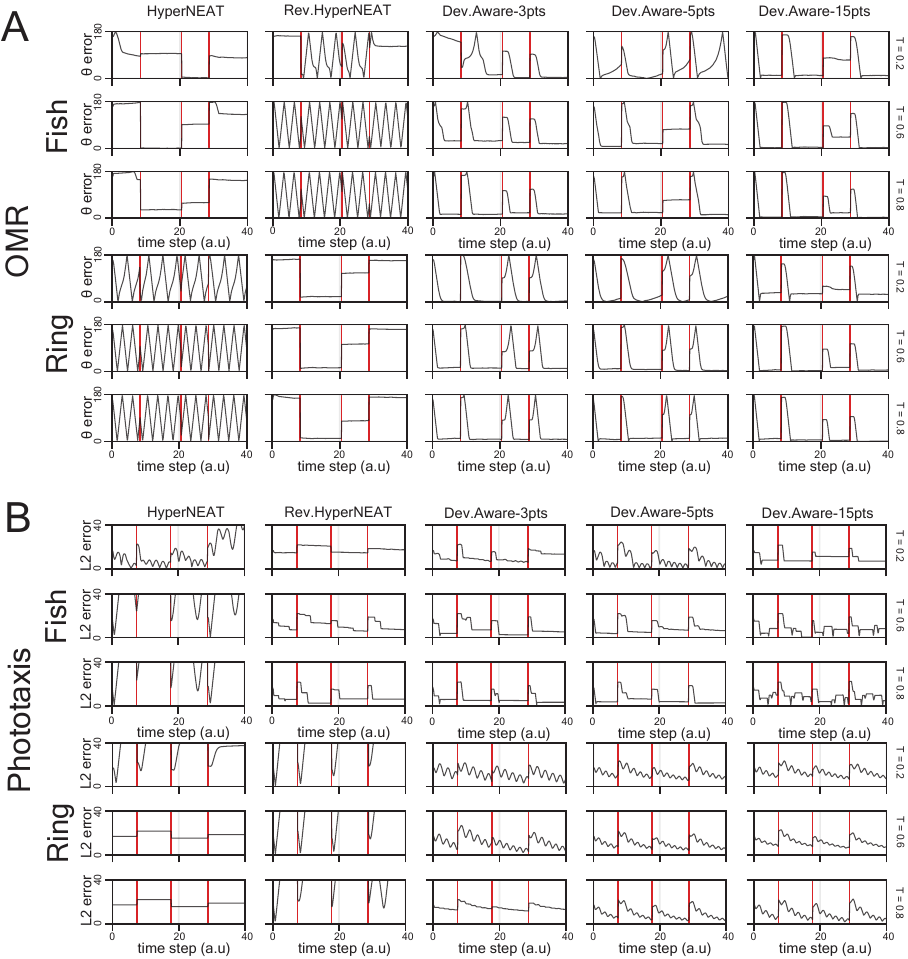}
\caption{Behavioral Stability Across Development for Best Seeds. (A) OMR performance during a fixed random trial, showing grating angle transitions (red) and error in desired orientation (black). Successful OMR drives orientation error toward zero. Flat lines indicate no change in orientation, sawtooth patterns indicate constant-rate spinning, and jumps at red lines reflect changes in grating angle. Although this metric cannot distinguish turning without movement, such behavior was never observed. Error trajectories are shown for three developmental times unseen during training (right). Despite learning OMR at their training times, HyperNEAT and Rev. HyperNEAT fail to turn in the correct direction at these unseen times. (B) Phototaxis performance during a fixed random trial, showing emitter position transitions (red) and the L2 distance between the agent and emitter center (black). Successful phototaxis drives this distance toward zero; jumps at red lines reflect changes in emitter position. As in OMR, HyperNEAT and Rev. HyperNEAT fail to generalize to the three unseen developmental times, with errors sometimes exceeding the plotted range. Oscillatory behavior, particularly on the Ring substrate, reflects circular movement of varying radii. We hypothesize that this strategy evolved because the emitter can move beyond the retina's detection range, favoring continuous environmental scanning for luminance changes.}\label{fig3}
\end{figure}

\subsection{A shared generative program produced changing connectomes across development}
We next visualized the connectomes produced by the training-selected CPPNs. Connectivity matrices at $T = 0.0$ and $T = 1.0$ revealed substantial changes in network size and connectivity patterning between the developmental endpoints for every substrate–task combination (Fig. 4, top). These changes reflected both the addition and movement of neurons specified by the substrate growth model and the connections generated by the CPPN for the substrate present at each age. The matrices therefore illustrate that functional stability did not require an invariant connectome. Rather, the same compact generator produced different connectomes as the available neurons and their geometry changed. However, coherent spatiotemporal structure could be seen in the connectomes across development (Fig. 4, bottom).

\begin{figure}[h]
\centering
\includegraphics[width=0.8\textwidth]{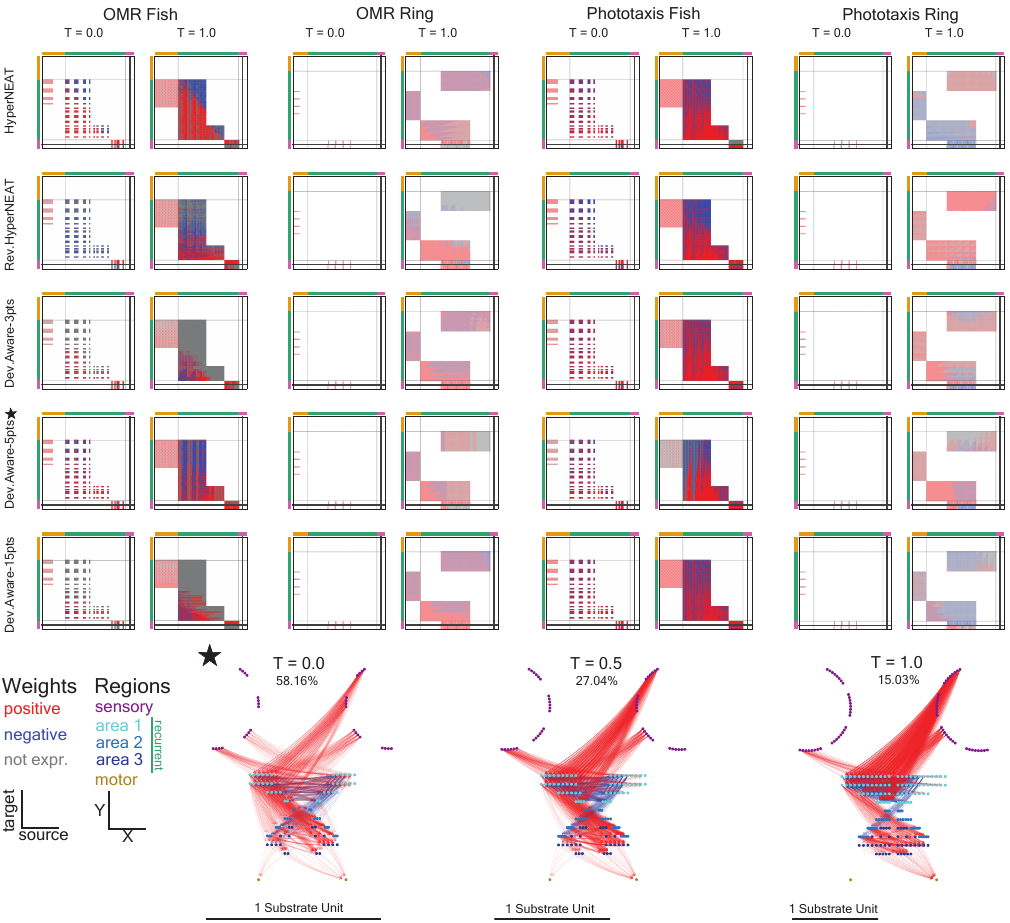}
\caption{Changes in Connectome Across Development. (Top) Connectivity matrices at the beginning and end of development for the best seeds across substrate–task pairs. Each entry represents the weight from a source (x-axis) to a target neuron (y-axis). Red and blue indicate positive (excitatory) and negative (inhibitory) weights respectively, gray indicates weights set to zero by the expression gate, and white indicates connections prohibited by projection constraints. Axis colors denote neuron locations in the substrate. Thin gray lines separate regions, thick gray lines separate the left and right motor neurons (enlarged relative to other neurons for ease of visualization). (Bottom) Connectome visualizations at three developmental snapshots for the OMR Fish Dev. Aware–5pts condition (star). Red and blue indicate positive and negative weights. Only the 2,000 connections with the largest absolute weights are shown per snapshot, with the percentage indicating the fraction of the full connectome displayed. Stronger inputs to the right motor neuron bias the agent toward rightward turns.}\label{fig4}
\end{figure}

\subsection{Stable behavior emerged through multiple connectome-level strategies}

We finally asked whether developmentally aware controllers shared a simple connectome-level strategy for maintaining function. For each best seed, we measured connection sparsity ratio, the proportion of expressed connections that were excitatory, and the mean weight of nonzero connections across development (Fig. 5A–D). Sparsity ratio was defined as the fraction of allowable connections assigned a weight of zero, including both zero-valued weight outputs and connections suppressed by the expression gate (i.e., a sparsity of 0.0 indicates all allowable weights were nonzero).

None of these coarse connectome statistics displayed a common developmental trajectory across the four substrate–task combinations or across the three developmentally aware conditions (Fig. 5A–D). Some controllers became more sparse, whereas others remained densely connected or changed at different rates. Excitatory connection ratios and mean nonzero weights similarly increased, decreased, or remained approximately constant depending on the evolved controller. For example, most selected controllers for phototaxis on the Fish substrate remained densely connected, while Dev. Aware–5pts became progressively more sparse. The selected OMR–Ring developmentally aware controllers also remained densely connected in most conditions, except for an increase in sparsity near the middle of development in Dev. Aware–5pts. In the phototaxis–Ring combination, the mean connection weight of Dev. Aware–15pts shifted from positive to negative over development (Fig. 5A–D, left). These observations do not exclude shared mechanisms at finer spatial or dynamical scales, but they show that developmental stability was not associated with a single pattern among the coarse connectome properties we measured.

The evolved CPPNs nevertheless almost always incorporated developmental time into their generated connectivity. Developmental time had an active path to the weight output, the expression output, or both in 58 of the 60 developmentally aware runs. The two exceptions occurred in phototaxis–Fish runs that were not selected as the highest-performing run for their respective conditions. All 12 developmentally aware best seeds structurally incorporated developmental time. Thus, although the resulting connectomes did not share a simple trajectory in sparsity, excitation, or mean weight, the generators almost always used developmental time as part of their connectivity rule.

Finally, we tested whether the controllers relied on neurons added during development. At each age, we calculated,
\begin{equation}
    \Delta F(T) = F_{\text{full growth}}(T) - F_{\text{no neurogenesis}}(T)
\end{equation}
where the no neurogenesis condition retained only the neurons present at $T=0$ while allowing neuronal position and intrinsic time constants to develop normally. Positive values therefore indicate that later-born neurons improved performance, negative values indicate that later-born neurons diminished performance, whereas values near zero indicate similar performance with and without those neurons (Fig. 5).

The meaning of a near-zero $\Delta F(T)$ depended on the controller’s absolute fitness. HyperNEAT controllers generally showed low fitness later in development under both full growth and no neurogenesis, resulting in values of $\Delta F(T)$ near zero (Figs. 2C,D and 5). Suppressing neurogenesis therefore did not rescue their developmental loss of function. This result indicates that neurogenesis was not the sole cause for their failure and that developmental changes retained in the no neurogenesis condition, including neuronal displacement and changes in intrinsic time constants, were sufficient to disrupt performance. Rev. HyperNEAT controllers instead generally showed positive values near $T = 1.0$, where the fully developed networks had been optimized and later-born neurons contributed to their mature performance (Fig. 5). Near-zero differences at earlier ages reflected similarly poor performance with and without neurogenesis rather than stable function.

Most developmentally aware controllers showed positive $\Delta F(T)$ during at least part of development while maintaining high absolute fitness, indicating that they incorporated later-born neurons into functional circuits (Figs. 2C,D and 5). However, the best seeds of the Dev. Aware–15pts controllers for both OMR substrates maintained high performance while exhibiting little difference between full development and the no-neurogenesis condition (Fig. 5A,B). These controllers therefore evolved strategies that were largely independent of later-born neurons, even though the full generated connectomes contained connections involving those neurons (Fig. 4). Developmentally stable behavior could consequently be achieved either by exploiting neurogenesis or by producing a controller whose function remained largely supported by the neurons present at the beginning of development. Together with the heterogeneous connectome statistics, these results indicate that the generative encoding discovered multiple substrate- and task-dependent solutions rather than converging on a single simple developmental strategy.

\begin{figure}[h]
\centering
\includegraphics[width=1.0\textwidth]{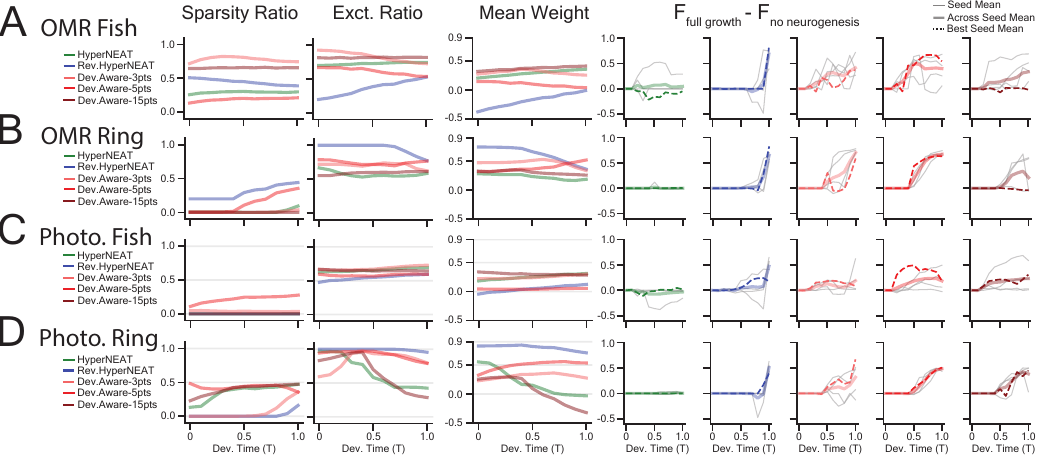}
\caption{Investigation of Developmental Strategies. (A) OMR–Fish pair. (Left) Developmental trajectories of connectome sparsity ratio, excitatory ratio, and mean nonzero connection weight. (Right) Fitness difference between the full growth model and a model without neurogenesis. Positive values indicate improved performance with neurogenesis, negative values indicate improved performance without it, and values near zero indicate little effect. All differences are zero at $T=0$, when neurogenesis has not yet occurred. Gray lines show individual seeds, thick transparent lines show condition means, and dotted lines indicate the best seed. (B–D) The same analyses for the remaining substrate–task pairs.}\label{fig5}
\end{figure}

\section{Methods}
\subsection{Developmental substrates}

Development was parameterized by a continuous variable $T \in [0,1]$, with $T=0$ denoting the initial substrate and $T=1$ the fully developed substrate. Both substrates were bilaterally symmetric and comprised sensory neurons receiving environmental observations, recurrent neurons organized into three areas with area-specific projection rules, and two motor neurons providing action outputs. The Fish geometry arranged the recurrent areas along an approximately rostrocaudal axis, whereas the Ring geometry arranged them around a peripheral structure with a small initial recurrent population. Although the Fish substrate was inspired by the zebrafish visuomotor circuit \citep{naumann2016whole}, both substrates were abstract computational models rather than reconstructions of an individual animal. Neuron identifiers were retained across developmental ages, allowing a fixed maximum state-vector size (the substrate at $T=1.0$), while an existence mask determined which neurons were present at each age. 

Each recurrent layer specified early, middle, and late neuron counts per hemisphere. Early neurons were present at $T=0$. For a middle or late cohort containing $n$ neurons and assigned a birth window $[a,b]$, birth times were deterministically assigned as

$$
T_k = a + \frac{k}{n+1}(b-a), \qquad k=1,\ldots,n.
$$

Births therefore occurred at equally spaced points strictly inside each window. This rule was applied separately to each recurrent layer and sensory population, with sensory counts specified per eye for bipolar cells and per group per eye for direction-selective ganglion cells (DSGCs). Corresponding bilateral neurons shared birth times, and a neuron was present whenever $T$ reached or exceeded its assigned birth time. Motor neurons were present throughout development. Birth-window bounds and other developmental parameters are given in Table 1.

Within each recurrent layer, neuron positions were assigned in early, middle, and late cohort order from the outer edge toward the midline in Fish and in the reverse direction in Ring. Because Fish's middle and late birth windows overlapped, this spatial ordering did not imply a strictly monotonic lateral sequence of births across cohorts. Ring began with 20 recurrent neurons confined to Area 2; Areas 1 and 3 initially contained none. Its middle cohorts contained two or three neurons per layer per hemisphere. With the middle birth window $[0.35,0.70]$, the first births in these cohorts occurred at $T\approx0.4667$ and $T=0.4375$, respectively. Thus, Ring underwent no additional recurrent neurogenesis before $T=0.4375$, although its initial recurrent population was already present.

Bilateral neuron positions were mirrored about the midline. Brain dimensions, eye positions and radii, and body length were interpolated between their initial and final values according to

$$
p(T)=p_0+(p_1-p_0)g(T),
\qquad
g(T)=
\begin{cases}
T, & \text{Fish},\\
T^2, & \text{Ring}.
\end{cases}
$$

This geometric growth interpolation was separate from the birth-time rule. The CTRNN controller time constant $\tau$ followed the same substrate-specific interpolation, giving $\tau(T)=0.4-0.2T$ for Fish and $\tau(T)=0.4-0.2T^2$ for Ring.

\begin{table}[h]
\caption{HyperNEAT substrate parameter values for the Fish and Ring substrates}\label{tab1}%
\begin{tabular}{@{}lll@{}}
\toprule
Parameter & Fish & Ring \\
\midrule
Developmental interpolation, $g(T)$ & $T$ & $T^2$ \\
Lateral assignment of recurrent cohort ranks & Outside to inside & Inside to outside \\
Middle-cohort birth-window bounds & $[0.00,0.60]$ & $[0.35,0.70]$ \\
Late-cohort birth-window bounds & $[0.40,1.00]$ & $[0.70,1.00]$ \\
Recurrent neurons, initial $\rightarrow$ final & $80\rightarrow164$ & $20\rightarrow296$ \\
Brain width, initial $\rightarrow$ final & $1.00\rightarrow2.00$ & $0.70\rightarrow1.80$ \\
Brain length, initial $\rightarrow$ final & $1.00\rightarrow2.00$ & $0.90\rightarrow1.80$ \\
Eye radius, initial $\rightarrow$ final & $0.25\rightarrow0.50$ & $0.10\rightarrow0.30$ \\
Field of view per eye; vergence parameter & $163^{\circ};\ 71.2^{\circ}$ & $140^{\circ};\ 0^{\circ}$ \\
Bipolar cells per eye, initial $\rightarrow$ final & $16\rightarrow32$ & $16\rightarrow32$ \\
Phototaxis sensing distance, initial $\rightarrow$ final & $20\rightarrow40$ & $20\rightarrow40$ \\
Samples per phototaxis ray, initial $\rightarrow$ final & $5\rightarrow10$ & $5\rightarrow10$ \\
DSGCs per eye, initial $\rightarrow$ final & $16\rightarrow32$ & $16\rightarrow32$ \\
Motor neurons throughout development & $2$ & $2$ \\
CTRNN $\tau$, initial $\rightarrow$ final & $0.4\rightarrow0.2$ & $0.4\rightarrow0.2$ \\
\botrule
\end{tabular}
\footnotetext{Recurrent counts include both hemispheres. Brain dimensions and eye radii are expressed in substrate coordinate units, phototaxis sensing distances in environment coordinate units, and $\tau$ in simulation time units. Phototaxis sampling parameters apply only to bipolar-like detectors; OMR input uses the directional signals defined above.}
\end{table}

\begin{figure}[h]
\centering
\includegraphics[width=0.8\textwidth]{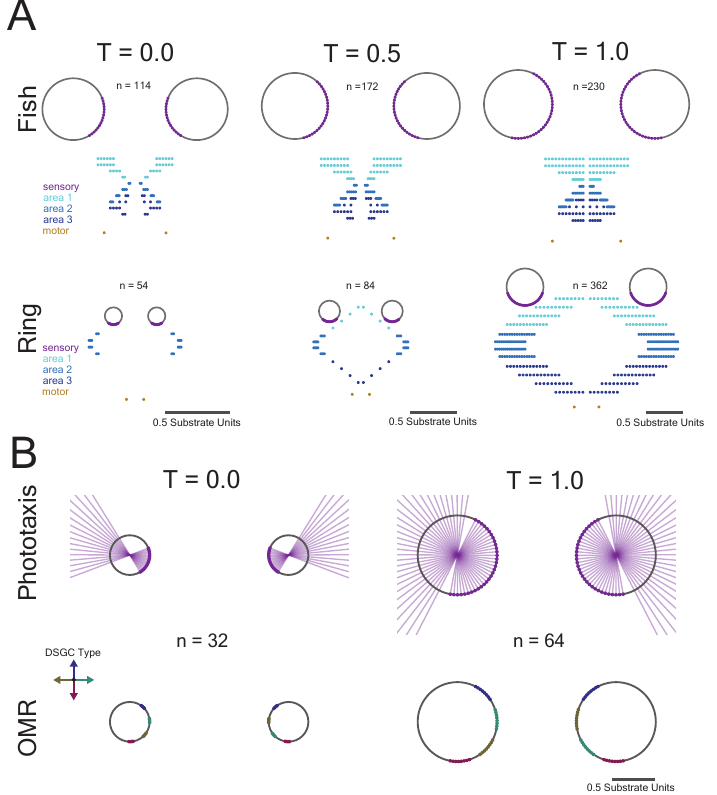}
\caption{Substrate Growth Models. (A) Three developmental snapshots (\(T\)) of the Fish and Ring substrates. Each differs in neuronal migration, neurogenesis schedules, and projection constraints. The number of neurons available to the CPPN is shown for each snapshot. Each substrate neuron corresponds to a CTRNN neuron, with motor neuron outputs determining the agent's actions (see Methods). (B) Retinal models used for phototaxis (top) and OMR (bottom). Bipolar cells detect luminance averaged across points along a projected ray (see Methods); their number and effective detection range double across development. Direction-selective ganglion cells (DSGCs) detect local optic flow from moving gratings. Both substrates use the same initial and final cell counts for each retinal model.}\label{fig6}
\end{figure}

\subsection{Projection constraints}
A directed source–target pair could receive a CPPN-generated weight only if its two neurons existed at the queried age and the pair matched a substrate-specific projection rule (Table 2). Projection rules defined allowable anatomical pathways whereas the CPPN expression gate determined which of those allowable pairs became nonzero connections. In Fish, outgoing sensory weights were converted to their absolute values, constraining them to be nonnegative. Other allowable weights could be positive or negative. No symmetry constraint was imposed on the resulting weights.

\begin{table}[h]
\caption{Allowed source $\rightarrow$ target projections for the Fish and Ring substrates}\label{tab2}%
\begin{tabular}{@{}lp{10cm}@{}}
\toprule
Substrate & Allowed source $\rightarrow$ target projections \\
\midrule
Fish & Sensory L $\rightarrow$ Area 1 R; sensory R $\rightarrow$ Area 1 L; Area 1 $\rightarrow$ Area 1, Area 2, and Area 3; Area 2 $\rightarrow$ Area 3; Area 3 $\rightarrow$ motor. \\[6pt]
Ring & Sensory L $\rightarrow$ Area 2 R; sensory R $\rightarrow$ Area 2 L; Area 2 R $\rightarrow$ motor L, Area 1 L, and Area 3 L; Area 2 L $\rightarrow$ motor R, Area 1 R, and Area 3 R; Area 1 L $\rightarrow$ Area 3 R; Area 1 R $\rightarrow$ Area 3 L; Area 3 L $\rightarrow$ Area 1 R; Area 3 R $\rightarrow$ Area 1 L. \\
\botrule
\end{tabular}
\footnotetext{Note that the Fish and Ring substrates had exclusively contralateral retinal projections, which are present in the zebrafish visuomotor system \citep{naumann2016whole}.}
\end{table}

\subsection{Sensory inputs and recurrent controller}

Each potential substrate neuron was assigned an entry in a fixed-size state vector, with an existence mask determining which neurons were present at developmental age $T$. Sensory states $\mathbf{s}$ acted as leaky filters of the external sensory inputs $\mathbf{I}(t)$:

$$
\tau(T)\frac{d\mathbf{s}}{dt}=-\mathbf{s}+\mathbf{I}(t).
$$

Here, $t$ denotes simulation time and $T$ denotes developmental age. Note that $T$ is held fixed during a trial.

Sensory populations expanded from central retinal positions toward their flanks, with corresponding cells mirrored between eyes. For phototaxis, each eye contained 16 early, 8 middle, and 8 late bipolar-like luminance detectors. Each active detector received the mean luminance at equally spaced sample points along a ray extending from its sensory location to the current sensing distance. The sensing distance increased from 20 to 40 environment units, and the number of samples increased from 5 to 10, following the substrate-specific interpolation rule (rounded to an integer).

For the optomotor response (OMR) task, each eye contained four spatially separated direction-selective ganglion cell (DSGC) groups with backward, leftward, rightward, and forward preferences. Each group contained four early, two middle, and two late cells. Early cells occupied the group core, and later cells were recruited on alternating flanks. Sensory input represented stimulus-motion direction relative to the agent's heading. Let $\phi$ denote the stimulus-motion direction in world coordinates and $\theta$ the agent's heading, both measured counterclockwise from the positive $x$-axis. The normalized motion vector's forward and leftward components in the agent's body frame were

$$
m_f=\cos(\phi-\theta),\qquad m_l=\sin(\phi-\theta).
$$

DSGC inputs were obtained by applying a ReLU function to these components:

$$
\begin{aligned}
I_{\mathrm{backward}} &= \max(0,-m_f), &
I_{\mathrm{forward}} &= \max(0,m_f),\\
I_{\mathrm{leftward}} &= \max(0,m_l), &
I_{\mathrm{rightward}} &= \max(0,-m_l).
\end{aligned}
$$

Every active cell sharing a directional preference received the same input in both eyes. These signals depended on stimulus direction and agent heading, without accounting for agent translation or estimating motion from local retinal image samples. Input magnitudes reflected directional alignment rather than stimulus speed. Retinal positions determined the cells' locations within the substrate but did not enter the directional-input calculation. Fish and Ring had identical sensory population sizes but differed in eye geometry and recurrent growth programs. Sensory growth parameters are given in Table 1.

The recurrent and motor states were collected in a vector $\mathbf{x}$. Their deterministic dynamics were

$$
\tau(T)\frac{d\mathbf{x}}{dt}
=-\mathbf{x}
+W_{\mathrm{sen}}(T)r(\mathbf{s})
+W_{\mathrm{rec}}(T)r(\mathbf{x}),
$$

where $W_{\mathrm{sen}}(T)$ maps sensory states to recurrent and motor neurons, and $W_{\mathrm{rec}}(T)$ contains the connections among non-sensory neurons. Recall that the goal of training is to find CPPNs that parameterize $W_{\mathrm{sen}}(T)$ and $W_{\mathrm{rec}}(T)$ such that the agent retains high fitness for any value of $T$. The activation function $r(z)$ of arbitrary variable $z$ was a clamped ReLU function, applied elementwise:

$$
r(z)=\min\{\max(z,0),10\}.
$$

Note, the controller contained no bias terms.

The dynamics were integrated using forward Euler with $\Delta t=0.1$ simulation time units. At each integration step, independent Gaussian perturbations were added to the input drives of recurrent and motor neurons. Thus, the non-sensory update was

$$
\mathbf{x}_{k+1}
=\mathbf{x}_k
+\frac{\Delta t}{\tau(T)}
\left[
-\mathbf{x}_k
+W_{\mathrm{sen}}(T)r(\mathbf{s}_k)
+W_{\mathrm{rec}}(T)r(\mathbf{x}_k)
+\boldsymbol{\eta}_k
\right],
$$

where the components of $\boldsymbol{\eta}_k$ were sampled independently across neurons and integration steps from $\mathcal{N}(0,0.01^2)$. Sensory states received no additive drive noise. At the beginning of each trial, all neuron states were independently initialized from $\mathcal{N}(0,0.01^2)$. After each integration step, states corresponding to neurons not yet born were masked (i.e., set to zero) and thus had no effect on the network activity. We emphasize that masking is mathematically equivalent to having CTRNNs of different topologies across $T$.

Noise was included during training to discourage controllers whose behavior depended on precisely unperturbed neural trajectories. In preliminary experiments without training noise, some controllers performed the tasks but exhibited large recurrent-state excursions when small perturbations were subsequently introduced. We hypothesize that the CPPNs obtained without noise produced ``hidden'' unstable dynamics, which went unperturbed during a trial. 

The two motor states were transformed into bounded left and right motor drives:

$$
d_L=\tanh(x_L),\qquad d_R=\tanh(x_R).
$$

Turn and forward commands were then calculated as

$$
\begin{aligned}
u_{\mathrm{turn}}
&=\operatorname{clip}\left(
\frac{d_L-d_R}{2}+\xi_{\mathrm{turn}},-1,1
\right),\\
u_{\mathrm{forward}}
&=\operatorname{clip}\left(
\frac{d_L+d_R}{2}+\xi_{\mathrm{forward}},0,1
\right),
\end{aligned}
$$

where $\xi_{\mathrm{turn}}$ and $\xi_{\mathrm{forward}}$ were independent samples from $\mathcal{N}(0,0.01^2)$, resampled at each step and independent of the neural-drive noise. These commands determined angular velocity and forward speed:

$$
\omega=\omega_{\max}u_{\mathrm{turn}},
\qquad
v=v_{\max}u_{\mathrm{forward}},
$$

with $\omega_{\max}=240^\circ$ per simulation time unit and $v_{\max}=30$ environment units per simulation time unit. At each step, the agent first updated its heading by $\omega\Delta t$ and then moved a distance $v\Delta t$ along the updated heading, subject to clipping at the environment boundaries. Agent movement had no inertia.

\subsection{Visuomotor tasks and fitness}
Agents moved within a two-dimensional environment measuring $1000\times1000$ environment units. Each episode began at the environment center with a heading sampled uniformly from $[-180^\circ,180^\circ)$. An episode comprised a sequence of stimulus blocks, with each block duration sampled independently and uniformly from the integers 60 through 140 simulation steps. With $\Delta t=0.1$, these durations corresponded to 6–14 simulation time units. Variable block durations reduced the predictability of stimulus transitions. Agent position, heading, and controller states continued across block transitions without being reset.

Each OMR episode contained eight blocks: two presentations of each of four stimulus-motion directions, corresponding to the world-coordinate vectors $(1,0)$, $(-1,0)$, $(0,1)$, and $(0,-1)$. The eight presentations were randomly permuted within each episode. Each phototaxis episode contained five blocks, with a newly sampled emitter position at the beginning of every block, including the first. Block sequences and durations were resampled each generation and shared by all genomes evaluated within that generation. For phototaxis, shared random draws specified emitter offsets relative to each agent's current position. Absolute emitter locations could therefore differ between agents following different trajectories.

For OMR, the target heading $\phi_b$ was the world direction of stimulus motion in block $b$. The reward at step $k$ was

$$
R^{\mathrm{OMR}}_{b,k}
=\cos(\theta_{b,k}-\phi_b)\,u_{\mathrm{forward},b,k},
$$

where $\theta_{b,k}$ was the agent's heading after the movement update and $u_{\mathrm{forward},b,k}\in[0,1]$ was the forward command defined in the controller description, including its motor noise and clipping. An aligned heading combined with a positive forward command received positive reward, a heading more than $90^\circ$ away from the target combined with a positive forward command received negative reward, and a zero forward command produced zero reward regardless of heading. The forward command was used instead of actual distance traveled to prevent the environment boundaries from penalizing otherwise appropriate OMR behavior. 

For phototaxis, a new emitter bearing was sampled uniformly from $[-\pi,\pi)$ and its distance from the agent's current position was sampled uniformly from 15 to 25 environment units at the beginning of each block. Sampled emitter coordinates were clipped to the environment boundaries. The emitter remained stationary within the block and produced a Gaussian luminance field,

$$
L(d)=\exp\left(-\frac{d^2}{2\sigma^2}\right),
\qquad \sigma=10,
$$

where $d$ was the distance from the emitter center in environment units. Peak intensity was 1.0 and ambient intensity was 0.0. The step reward was the luminance evaluated at the agent's center after movement:

$$
R^{\mathrm{photo}}_{b,k}
=L\left(\left\|\mathbf{p}_{b,k}-\mathbf{e}_b\right\|\right),
$$

where $\mathbf{p}_{b,k}$ was the agent's position and $\mathbf{e}_b$ was the emitter position. Thus, phototaxis reward depended on the luminance at the agent's center.

For either task, step rewards were averaged within each block and block scores were then averaged to obtain episode fitness:

$$
F_{\mathrm{episode}}
=\frac{1}{B}\sum_{b=1}^{B}
\left(\frac{1}{K_b}\sum_{k=1}^{K_b}R_{b,k}\right),
$$

where $B$ was the number of stimulus blocks and $K_b$ was the number of steps in block $b$. Each block therefore contributed equally to episode fitness regardless of its duration. Episode scores were averaged at each evaluated developmental age, and the resulting age-specific scores were averaged to obtain genome fitness.

\subsection{CPPN encoding and neuroevolution}
A compositional pattern-producing network (CPPN) encoded the controller's connectivity as a function of substrate geometry and developmental age. For each directed source–target pair permitted by the projection rules, the CPPN received seven inputs:

$$
\mathbf{q}_{ij}(T)
=\left(x_{\text{src}},y_{\text{src}},x_{\text{tgt}},y_{\text{tgt}},T,\frac{x_{\text{src}}-x_{\text{tgt}}}{2},\frac{y_{\text{src}}-y_{\text{tgt}}}{2}\right),
$$

where $(x_{\text{src}},y_{\text{src}})$ and $(x_{\text{tgt}},y_{\text{tgt}})$ were the source and target coordinates at the queried developmental age $T$. Fish and Ring used substrate coordinates directly, without coordinate normalization. The relative offsets were half the source-minus-target coordinate differences.

The CPPN had two outputs: a connection weight $w_{ij}$ and an expression gate $e_{ij}$. This formulation of HyperNEAT is referred to as HyperNEAT-LEO \citep{verbancsics2011constraining} and has been found to promote modularity in substrate network connectivity. Both output nodes used $\tanh$ activation. A connection was expressed only when $e_{ij}>0$:

$$
\widetilde{w}_{ij}(T)=
\begin{cases}
w_{ij}(T), & e_{ij}(T)>0,\\
0, & e_{ij}(T)\leq0.
\end{cases}
$$

The resulting weights were then subject to the sensory-sign and neuron-existence constraints. In Fish, outgoing sensory weights were made non-negative by taking their absolute values whereas Ring imposed no sensory-sign constraint. Connections involving a neuron not yet born were set to zero. The $\tanh$ weight output bounded all controller connection weights to $[-1,1]$, with Fish's outgoing sensory weights further restricted to $[0,1]$.

CPPNs were evolved using NEAT-style parameter and topology mutations, innovation-aligned crossover, speciation, and elitist selection. Initial CPPNs contained seven input nodes (geometric inputs above), two hidden nodes, and two output nodes (weight and expression gate outputs). Initial hidden-node activations were sampled uniformly from Gaussian, sigmoid, $\tanh$, $\cos$, $\sin$, step, and ReLU functions. CPPN connection parameters were initialized from $\mathcal{N}(0,1)$, and hidden- and output-node biases were initialized from $\mathcal{N}(0,0.1^2)$. Initial connections were constructed between input and hidden nodes, hidden and output nodes, and directly between input and output nodes. Connections were sampled with probability 0.7, with additional connections inserted to ensure that each hidden node had an incoming connection and a connection to each output. Each constructed connection was enabled independently with probability 0.8.

Mutation could perturb or reset CPPN connection parameters, perturb non-input-node biases, change hidden-node activations, enable or disable connections, add a feedforward connection, or split an enabled connection by inserting a hidden node. Inserted nodes initially used identity activation and zero bias. The original connection was disabled, the source-to-new-node connection received weight one, and the new-node-to-target connection inherited the original weight. This preserved the original connection's contribution when the split was introduced. All CPPN connections respected a feedforward node ordering.

Crossover aligned connection genes by innovation number. Matching genes were sampled from either parent, while disjoint and excess genes were inherited from the fitter parent. Genomes were grouped into species using a compatibility distance based on normalized excess and disjoint gene counts and the mean absolute weight difference between matching connection genes. Parents were selected by within-species tournaments. Global elites were retained unchanged, and remaining offspring were allocated among species using species mean fitness, with a non-negative shift when needed and one offspring reserved per species when capacity allowed. Crossover occurred with probability 0.4 for offspring from species containing at least two members, otherwise reproduction used a single parent. All non-elite offspring underwent mutation. The configured encoding and evolutionary parameters are summarized in Table 3.

\begin{table}[h]
\caption{CPPN encoding and neuroevolution parameters}\label{tab4}%
\begin{tabular}{@{}lp{8cm}@{}}
\toprule
Parameter & Value or rule \\
\midrule
Initial input / hidden / output nodes & 7 / 2 / 2 \\
Weight-output activation & $\tanh$ \\
Expression-output activation & $\tanh$ \\
Connection expression criterion & Expression output $>0$ \\
Decoded controller weight bounds & $[-1,1]$; outgoing sensory weights in Fish: $[0,1]$ \\
Initial and mutated hidden-activation choices & Gaussian, sigmoid, $\tanh$, $\cos$, $\sin$, step, ReLU \\
Initial CPPN connection-parameter distribution & $\mathcal{N}(0,1)$ \\
Initial hidden- and output-node bias distribution & $\mathcal{N}(0,0.1^2)$ \\
Initial connection sampling probability & 0.70, subject to the construction rules described above \\
Initial connection-enabled probability & 0.80 per constructed connection \\
Connection-parameter mutation probability & 0.25 per connection gene \\
Reset probability, conditional on parameter mutation & 0.10; resample from $\mathcal{N}(0,1)$ \\
Otherwise, additive parameter perturbation & $\mathcal{N}(0,0.1^2)$ \\
Disable-connection mutation probability & 0.05 per offspring; at most one enabled connection disabled \\
Enable-connection mutation probability & 0.05 per offspring; at most one disabled connection enabled \\
Add-connection mutation probability & 0.10 per offspring \\
Add-node mutation probability & 0.10 per offspring \\
Inserted hidden-node activation and bias & Identity; zero \\
Hidden-activation mutation probability & 0.05 per hidden node \\
Bias mutation probability & 0.05 per non-input node \\
Additive bias perturbation & $\mathcal{N}(0,0.5^2)$ \\
Crossover probability & 0.40, when the species contains at least two members \\
Disabled-gene inheritance probability & 0.75 for matching genes disabled in either parent \\
Parent selection & Within-species tournament; configured tournament size 5 \\
Elitism & Global highest-fitness genomes retained unchanged \\
Compatibility threshold & 1.0 \\
Excess / disjoint / matching-weight coefficients & 1.0 / 1.0 / 0.4 \\
Compatibility normalization & Maximum connection-gene count of the two genomes, with a minimum of 1 \\
\botrule
\end{tabular}
\end{table}

\subsection{Training conditions and developmental testing}

We crossed two substrates (Fish and Ring) with two visuomotor tasks (OMR and phototaxis) and compared five training conditions within each substrate–task pairing. HyperNEAT was optimized exclusively on the initial substrate ($T=0$), whereas Rev. HyperNEAT was optimized exclusively on the mature substrate ($T=1$). Three developmentally aware conditions were optimized at 3, 5, or 15 ages spanning $T=0$ to $T=1$. The three-age schedule used $T=0,0.5,1$, and the five-age schedule used $T=0,0.25,0.5,0.75,1$. The 15-age schedule used approximately equally spaced ages, given by $k/14$ for $k=0,\ldots,14$, rounded to two decimal places. Developmental age remained fixed within each episode and training across ages evaluated the same CPPN on separately constructed developmental substrates.

All conditions used 15 evaluation episodes per genome per generation, divided equally among their training ages. Each endpoint-trained condition used 15 episodes at its single training age. The developmentally aware conditions used five, three, or one episode per age for the 3-, 5-, and 15-age schedules, respectively. Genome fitness was calculated by averaging episode fitness at each age and then averaging the age-specific scores:

$$
F_{\mathrm{genome}}
=\frac{1}{A}\sum_{a=1}^{A}
\left(\frac{1}{E}\sum_{j=1}^{E}F_{a,j}\right),
\qquad A \times E=15,
$$

where $A$ was the number of training ages, $E$ was the number of episodes per age, and $F_{a,j}$ was the episode fitness defined above. The conditions therefore had equal episode budgets but differed in developmental coverage and the number of episodes sampled at each age. Equal episode counts did not imply equal computational cost, because controller size varied with developmental age and block durations were sampled independently. Five independent evolutionary runs were performed for each training condition and substrate–task pairing, using populations of 100 genomes evolved for 200 generations.

For each training condition within each substrate–task pairing, the run selected for ``best seed" analysis was the one with the highest mean training fitness over the final 10 generations. Best seed selection used training performance only, without reference to subsequent testing performance. Evolved controllers were then evaluated without further evolution at 11 equally spaced developmental ages,

$$
T_{\mathrm{test}}\in\{0,0.1,0.2,\ldots,1\},
$$

using 20 testing trials per age. These evaluations assessed performance across development, including ages absent from the corresponding training schedule. We also examined trajectories under fixed stimulus sequences at untrained ages, visualizing heading error relative to the stimulus-motion direction for OMR and distance from the emitter for phototaxis.

Testing distinguished the developmental age of the substrate from the explicit time input supplied to the CPPN. For endpoint-trained controllers, the CPPN time input remained fixed at its training value: $T=0$ for HyperNEAT and $T=1$ for Rev. HyperNEAT. Substrate geometry, neuron availability, and age-dependent controller parameters followed the testing age. Controller weights were regenerated from the coordinates of the testing substrate while retaining the evolved CPPN parameters. Thus, endpoint transfer assessed how the learned connectivity rule generalized across substrate growth without changing its explicit developmental-time input. For developmentally aware controllers, both the substrate age and the CPPN time input were set to the testing age.

\subsection{Connectome and neurogenesis analyses}

At each analysis age, connectome statistics were calculated over allowable directed source–target pairs that existed at the queried $T$ and were permitted by the projection rules. Pairs prohibited by the projection rules and pairs involving neurons not yet born were excluded. Allowable pairs with zero effective weight remained in the analysis, including zeros produced by the CPPN expression gate and zeros returned by the weight output itself. We used the effective controller weights after application of the expression gate and sensory-sign constraints.

Let $N_{\mathrm{allowable}}$ denote the number of allowable pairs, $N_{\mathrm{zero}}$ the number with zero effective weight, and $N_{\mathrm{positive}}$ the number with positive effective weight. The sparsity ratio was defined as
$$
S=\frac{N_{\mathrm{zero}}}{N_{\mathrm{allowable}}}.
$$

We also calculated the positive-connection fraction among nonzero weights,

$$
P_+=\frac{N_{\mathrm{positive}}}{N_{\mathrm{nonzero}}},
$$

and the mean signed nonzero weight,

$$
\overline{w}_{\mathrm{nonzero}}
=\frac{1}{N_{\mathrm{nonzero}}}
\sum_{(i,j):\,w_{ij}\neq0}w_{ij},
$$

where the sum was restricted to allowable pairs. Zero weights were identified by exact comparison with zero. Both $P_+$ and $\overline{w}_{\mathrm{nonzero}}$ assessed different notions of excitatory and inhibitory balance of the controllers. Values near $1$ for $P_+$ indicated there were more excitatory synapses than inhibitory ones, and positive values of $\overline{w}_{\mathrm{nonzero}}$ indicated that there was more excitatory input in the network in general.

To examine whether the CPPN could transmit its explicit developmental-time input to its outputs, we tested directed graph reachability from the $T$ input node to the weight and expression output nodes separately. Paths were constructed from enabled CPPN connection genes, including enabled genes with zero weight. As stated in Results, 58 out of 60 CPPNs had a valid path between $T$ and at least one of the output nodes.

To assess the contribution of later-born neurons, we compared each evolved controller under full growth with an intervention that retained only neurons present at $T=0$. At each testing age, the restricted network excluded all later-born sensory and recurrent neurons and their incident connections. The retained neurons occupied their prescribed positions at the testing age, and eye geometry, phototaxis sampling parameters, and the controller time constant continued to follow their developmental schedules. The intervention therefore restricted neuron recruitment while preserving the prescribed growth of the remaining substrate. The CPPN genome was unchanged, and neither condition involved retraining.

Full and restricted networks were evaluated using corresponding testing seeds. At each age, we calculated the mean paired fitness difference,

$$
\Delta F(T)
=\frac{1}{K}\sum_{j=1}^{K}
\left[F_{\mathrm{full},j}(T)-F_{\mathrm{restricted},j}(T)\right],
$$

where $K$ was the number of testing trials. Positive differences indicated that including later-born neurons improved performance in the evolved controller under the tested conditions and negative differences indicated better performance after their exclusion. This comparison assessed the combined contribution of later-born sensory and recurrent neurons and their connections.

\section{Discussion}\label{sec12}
Neural development requires a controller to remain functional while the network performing the computation changes. In our models, development repeatedly altered which neurons were present and which connections were possible, while neuronal positions and intrinsic time constants also changed with age. Despite these changes in topology, geometry, and dynamics, a single evolved compositional pattern-producing network generated connectivity that maintained visuomotor performance across the developmental trajectory. Controllers optimized at one endpoint generally failed as the substrate moved away from that age. In contrast, controllers optimized across development maintained performance at trained and untrained ages in two distinct substrates and two behaviors. In some conditions, only three developmental snapshots were required during optimization.

Animals preserve useful behavior while their nervous systems grow and remodel, but artificial neural networks have rarely been studied under the same requirement. Most developmental encodings evaluate the controller produced after growth. Previous HyperNEAT work instead showed that a common connectivity rule could be applied at increasing substrate densities \citep{stanley2009hypercube}, while \citep{risi2013confronting} generated controllers for changing body morphologies while keeping the neural substrate fixed. Here, the changing object was the controller itself. Its neuronal composition, geometry, available connectivity, and intrinsic dynamics varied across age. To our knowledge, this study is among the first to require an artificial recurrent controller to preserve a selected behavior across a developmental trajectory that combines these forms of neural change.

Performance at untrained ages suggests that the CPPNs captured regularities shared across developmental stages rather than requiring an independently optimized connectome at every age. The results also show that stability did not require an invariant connectome or a single trajectory of sparsity, weight sign, or mean connection strength. Some controllers benefited from later-born neurons, whereas others remained effective when those neurons were removed. Stable behavior could therefore arise through different developmental connectivity programs. Biological circuits can likewise produce similar output through different parameter configurations \citep{prinz2004similar}. Developing circuits can also recruit different homeostatic mechanisms across ages and perturbations \citep{tien2018homeostatic}, while growth can preserve motor output without identical juvenile and adult parameters \citep{bucher2005animal}. Our controllers do not reproduce those biological mechanisms, but they demonstrate that behavioral stability need not depend on one universal pattern of connectivity change.

Several limitations define the scope of this result. The substrate developmental trajectories were prescribed rather than evolved, so the study shows that generated connectivity can compensate for neural development but does not explain how the trajectory itself arises. Both substrates were abstract computational models rather than biologically grounded brain structures. The Fish substrate was inspired by zebrafish visuomotor organization but was not a reconstruction of an individual animal, and the modeled OMR used abstract directional signals rather than retinal images and locally computed optic flow. Additionally, the Ring substrate was not modeled after any particular animal, but chosen to show an extreme form of neural development. The results establish a computational principle rather than a detailed model of a particular animal's development.

Body mechanics and body growth were held fixed so that the experiments isolated neural substrate development. Nervous systems and bodies develop together, and changing biomechanics may require different controllers across age. This extension appears compatible with the present approach. \citet{risi2013confronting} showed that morphology-conditioned HyperNEAT can generate controllers across changes in quadruped leg length and interpolate to unseen body morphologies. Neural age and body morphology could therefore be supplied jointly to a connectivity generator. We considered this combined problem outside the scope of the current study. More generally, functional stability is relative to a selected behavior. Natural development may preserve some sensorimotor functions while appropriately changing others as the body and behavioral demands change.

One extension would allow the developmental program and functional controller to evolve together. Several existing approaches provide components for such a model. ES-HyperNEAT uses CPPNs to determine hidden-neuron placement and density rather than operating on a fully specified substrate \citep{risi2012enhanced}. Neural developmental programs instead evolve local rules that add neurons, differentiate cell states, and construct network topology during growth \citep{najarro2023towards}. Related work has examined the role of neuronal diversity in growing controllers \citep{nisioti2024growing} and extended self-assembling networks to structural change and experience-dependent learning during an agent’s lifetime \citep{plantec2024evolving}. These methods demonstrate that neuronal composition and connectivity can emerge from an evolved developmental process, but they generally optimize the controller produced by development or its ability to adapt rather than requiring a selected behavior to remain effective throughout development. Combining these developmental encodings with the trajectory-level objective used here would allow neuronal birth, position, intrinsic properties, and connectivity to co-evolve under selection for function at every developmental stage. A successful mature controller would not be sufficient if its intermediate forms failed to behave effectively. Different encodings could represent development through global geometric rules, local cell-level interactions, or a combination of the two. 

The framework also supports a broader research program on how developmental structure affects the ability to learn stable behavior. HyperNEAT is sensitive to the geometric representation supplied to the CPPN \citep{clune2009sensitivity}, and the Fish and Ring substrates produced different optimization variability and connectivity trajectories. Systematically varying substrate geometry, neuronal birth order, migration, projection constraints, and rates of intrinsic change could reveal which developmental organizations facilitate functional stability. Training across families of growth models could further test whether a connectivity program generalizes to individual differences in developmental timing.

A complementary direction would prescribe development from biological measurements and evaluate the resulting connectivity programs in embodied models of well-characterized animals. Existing neuromechanical models have combined biologically derived neural controllers with anatomically informed bodies, sensory systems, biomechanics, and closed-loop interactions with the environment \citep{lobato2022neuromechfly,liu2025artificial}. Model organisms such as zebrafish and Drosophila are particularly promising because extensive experimental data describe their neural organization, behavior, body mechanics, and development. Age-dependent neuronal populations, positions, projection constraints, sensory organization, intrinsic properties, and body dimensions could be incorporated into these embodied platforms. Evolution could then infer connectivity rules that preserve behavior across the measured developmental trajectory. This would allow developmental connectivity programs to be evaluated within a biologically constrained brain, body, and environment system rather than in an abstract controller alone. Such models could predict which neuronal populations are necessary at different ages, which connections must change to compensate for neural and bodily growth, and which developmental perturbations should produce age-specific behavioral failures. While such models would be computationally demanding, there have been recent advances in accelerating the speed and efficiency of evolutionary algorithms. EvoJAX places the evolutionary algorithm, network, and task on GPUs or TPUs \citep{tang2022evojax} whereas TensorNEAT directly supports accelerated NEAT, CPPN, and HyperNEAT populations \citep{wang2026tensorneat}. \citet{sarkar2025evolution} have further demonstrated evolutionary optimization of billion-parameter networks using low-rank perturbations. These methods address different computational bottlenecks, but together they suggest that evolutionary optimization need not remain restricted to small substrates.

\section{Conclusion}\label{sec13}
We showed that a compact connectivity generator conditioned on developmental time can preserve visuomotor function while a recurrent neural substrate undergoes repeated topological changes and continuous changes in geometry and intrinsic dynamics. Optimization at only a few developmental snapshots was often sufficient to maintain performance at untrained ages. Stability did not require an invariant connectome or a single connectivity-level strategy. These results extend generative neuroevolution from density scaling and morphology-conditioned control to functional connectivity across changing neural substrates. The framework provides a basis for studying artificial neural networks as developing systems and for asking which growth programs, substrate organizations, and biological constraints allow behavior to remain effective during neural development.

\backmatter


\bmhead{Acknowledgements}

Research reported in this publication was supported by the National Institutes of Health under Award Number R01 EY033845–01. The content is solely the authors’ responsibility and does not necessarily represent the official views of the National Institutes of Health. J.J.M. was supported by a Ruth K. Broad Foundation Postdoctoral Fellowship. E.A.N. was supported by the Whitehall and Alfred P. Sloan Foundation.

\section*{Data and Code Availability}
All data and code used for this project will be publicly released upon publication.


\bibliography{sn-bibliography}

\end{document}